\documentclass[RNAAS]{aastex62}

\usepackage{rotating} 

\graphicspath{{./}{figures/}}

\begin{document}

\title{Catalog of New K2 Exoplanet Candidates from Citizen Scientists}

\correspondingauthor{Jon Zink}
\email{jzink@astro.ucla.edu}

\author[0000-0003-1848-2063]{Jon K. Zink}
\affiliation{Department of Physics and Astronomy, University of California, Los Angeles, CA 90095}
\affiliation{Caltech/IPAC-NASA Exoplanet Science Institute, Pasadena, CA 91125}

\author[0000-0003-3702-0382]{Kevin K. Hardegree-Ullman}
\affiliation{Caltech/IPAC-NASA Exoplanet Science Institute, Pasadena, CA 91125}https://www.overleaf.com/23750108cfjmqhyykmmr

\author[0000-0002-8035-4778]{Jessie L. Christiansen}
\affiliation{Caltech/IPAC-NASA Exoplanet Science Institute, Pasadena, CA 91125}

\author[0000-0002-1835-1891]{Ian J. M. Crossfield}
\affiliation{Department of Physics, and Kavli Institute for Astrophysics and Space Research, Massachusetts Institute of Technology, Cambridge, MA 02139}

\author[0000-0003-0967-2893]{Erik A. Petigura}
\affiliation{Cahill Center for Astrophysics, California Institute of Technology, Pasadena, CA 91125, USA}
\affiliation{Sagan Fellow}

\author[0000-0001-5578-359X]{Chris J. Lintott}
\affiliation{Oxford Astrophysics, The Denys Wilkinson Building, Oxford OX1 3RH, United Kingdom}

\author[0000-0002-4881-3620]{John H. Livingston}
\affiliation{Department of Astronomy, University of Tokyo, 7-3-1 Hongo, Bunkyo-ku, Tokyo 113-0033, Japan}

\author[0000-0002-5741-3047]{David R. Ciardi}
\affiliation{Caltech/IPAC-NASA Exoplanet Science Institute, Pasadena, CA 91125}

\author[0000-0002-3306-3484]{Geert Barentsen}
\affiliation{Bay Area Environmental Research, Moffett Field, CA 94035}

\author[0000-0001-8189-0233]{Courtney D. Dressing}
\affiliation{Department of Astronomy, University of California, Berkeley 94720}

\author{Alexander Ye}
\affiliation{Department of Astronomy, University of California, Berkeley 94720}

\author[0000-0001-5347-7062]{Joshua E. Schlieder}
\affiliation{Exoplanets and Stellar Astrophysics Laboratory, Code 667, NASA Goddard Space Flight Center, Greenbelt, MD 20771, USA}

\author{Kevin Acres}
\affiliation{Exoplanet Explorers, Citizen Scientist}

\author{Peter Ansorge}
\affiliation{Exoplanet Explorers, Citizen Scientist}

\author{Dario Arienti}
\affiliation{Exoplanet Explorers, Citizen Scientist}

\author{Elisabeth Baeten}
\affiliation{Exoplanet Explorers, Citizen Scientist}

\author{Victoriano Canales Cerdá}
\affiliation{Exoplanet Explorers, Citizen Scientist}

\author{Itayi Chitsiga}
\affiliation{Exoplanet Explorers, Citizen Scientist}

\author{Maxwell Daly}
\affiliation{Exoplanet Explorers, Citizen Scientist}

\author{James Damboiu}
\affiliation{Exoplanet Explorers, Citizen Scientist}

\author{Martin Ende}
\affiliation{Exoplanet Explorers, Citizen Scientist}

\author{Adnan Erdag}
\affiliation{Exoplanet Explorers, Citizen Scientist}

\author{Stiliyan Evstatiev}
\affiliation{Exoplanet Explorers, Citizen Scientist}

\author{Joseph Henderson}
\affiliation{Exoplanet Explorers, Citizen Scientist}

\author{David Hine}
\affiliation{Exoplanet Explorers, Citizen Scientist}

\author{Tony Hoffman}
\affiliation{Exoplanet Explorers, Citizen Scientist}

\author{Emmanuel Lambrou}
\affiliation{Exoplanet Explorers, Citizen Scientist}

\author{Gabriel Murawski}
\affiliation{Exoplanet Explorers, Citizen Scientist}

\author{Mark Nicholson}
\affiliation{Exoplanet Explorers, Citizen Scientist}

\author{Mason Russell}
\affiliation{Exoplanet Explorers, Citizen Scientist}

\author{Hans Martin Schwengeler}
\affiliation{Exoplanet Explorers, Citizen Scientist}

\author{Alton Spencer}
\affiliation{Exoplanet Explorers, Citizen Scientist}

\author{Aaron Tagliabue}
\affiliation{Exoplanet Explorers, Citizen Scientist}

\author{Christopher Tanner}
\affiliation{Exoplanet Explorers, Citizen Scientist}

\author[0000-0001-5284-9231]{Melina Th\'evenot}
\affiliation{Exoplanet Explorers, Citizen Scientist}

\author{Christine Unsworth}
\affiliation{Exoplanet Explorers, Citizen Scientist}

\author[0000-0003-0788-8391]{Jouni Uusi-Simola}
\affiliation{Exoplanet Explorers, Citizen Scientist}



\section{Exoplanet Explorers} \label{sec:exoexp}

The \emph{K2} mission has successfully found $\approx\!1,000$ new exoplanet candidates.\footnote{\url{https://exoplanetarchive.ipac.caltech.edu}} Now with an enormous data set ($\approx~400,000$ stellar targets) that nearly doubles the source count of \emph{Kepler} \citep{hub16}, data parsing provides a unique time intensive obstacle. The Exoplanet Explorers\footnote{\url{https://www.zooniverse.org/projects/ianc2/exoplanet-explorers}} project, part of the Zooniverse platform, allows citizen scientists to help overcome the abundance of transit data \citep{chr18}. We make available 204,855 statistically significant dips in \emph{K2} light curves from campaigns 0-8, 10, and 12-14. We used the {\tt k2phot} pipeline \citep{Petigura18a} to remove the \emph{K2} systematics and searched for periodic transits using the {\tt TERRA} search algorithm \citep{Petigura13b}.  For training, each participant is shown an example of a real folded exoplanet transit light curve, with the expected model plotted over the data. The volunteer is then instructed to look for dips that provide a similar match to this basic transit model. Each folded light curve presented are assigned a ``Yes'' or ``No'' value by the citizen scientist, indicating their belief that the source of the dip is caused by a transiting exoplanet. This simple visual inspection helps create a targeted search of the \emph{K2} light curves.

\section{Criteria for Selection}\label{sec:crit}

For selection within this catalog we require additional cuts to provide a more focused list of potential candidates. We only consider dips that have received a ``Yes'' vote by $\ge90\%$ of the reviewers. All sources have been reviewed by at least 20 citizen scientists. Furthermore, we remove planet candidates that have been previously noted on the ExoFOP\footnote{\url{https://exofop.ipac.caltech.edu/k2/}} website (as of February 20, 2019). To eliminate eclipsing binaries, we exclude candidates with an inferred planet radius $R_{\rm pl}$ $>0.2R_{\Sun}$. An expert visual inspection of the fitted light curve is performed to further eliminate noisy and problematic fits. Strong V-shape transits have been removed. These grazing transits only provide a lower limit for the planetary radius, making it difficult to eliminate eclipsing binaries.

\section{Fit Method}\label{sec:fit}

Utilizing the {\tt EMCEE} algorithm \citep{for13}, we maximize and sample the posterior of the 8 transit parameters. The fitted parameters are as follows: $R_{\rm pl}/R_{\star}$ (the radius ratio of the planet and star), planet period, $T_0$ (the center of the first detected transit), $a_{\rm pl}/R_{\star}$ (the ratio of the semi-major axis to stellar radius), $b$ (the impact parameter), two quadratic limb darkening parameters, and a floating flux normalization parameter. Each parameter is fit using a uniform prior, with the exception of the semi-major axis ratio ($a_{\rm pl}/R_{\star}$) and the two limb darkening parameters. Assuming a perfect initial measurement of period from the from the {\tt TERRA} grid search, Kepler's law is used to derive a Gaussian prior for $a_{\rm pl}/R_{\star}$. The independent information of stellar radius and stellar mass are provided by \citet{hub16} with \emph{Gaia} radius updates from \citet{bai18}. Since the uncertainty in mass and radius are non-negligible, this constraint is rather weak. The priors for the limb darkening parameters are calculated using a Monte Carlo interpolation of the appropriate \citet{cla12} table. \

We use the {\tt BATMAN} transit model \citep{kre15} to fit the processed {\tt K2Phot} light curves. Here a Gaussian likelihood function is implemented to fit the photometric data to the transit model. This transit fitting procedure is similar to that of \citet{cro16}. The resulting parameters are provided in our supplementary \href{http://www.jonzink.com/blogEE.html}{blog}.

\section{Results}\label{sec:result}

\begin{figure}
\begin{center}
\includegraphics[width=\textwidth]{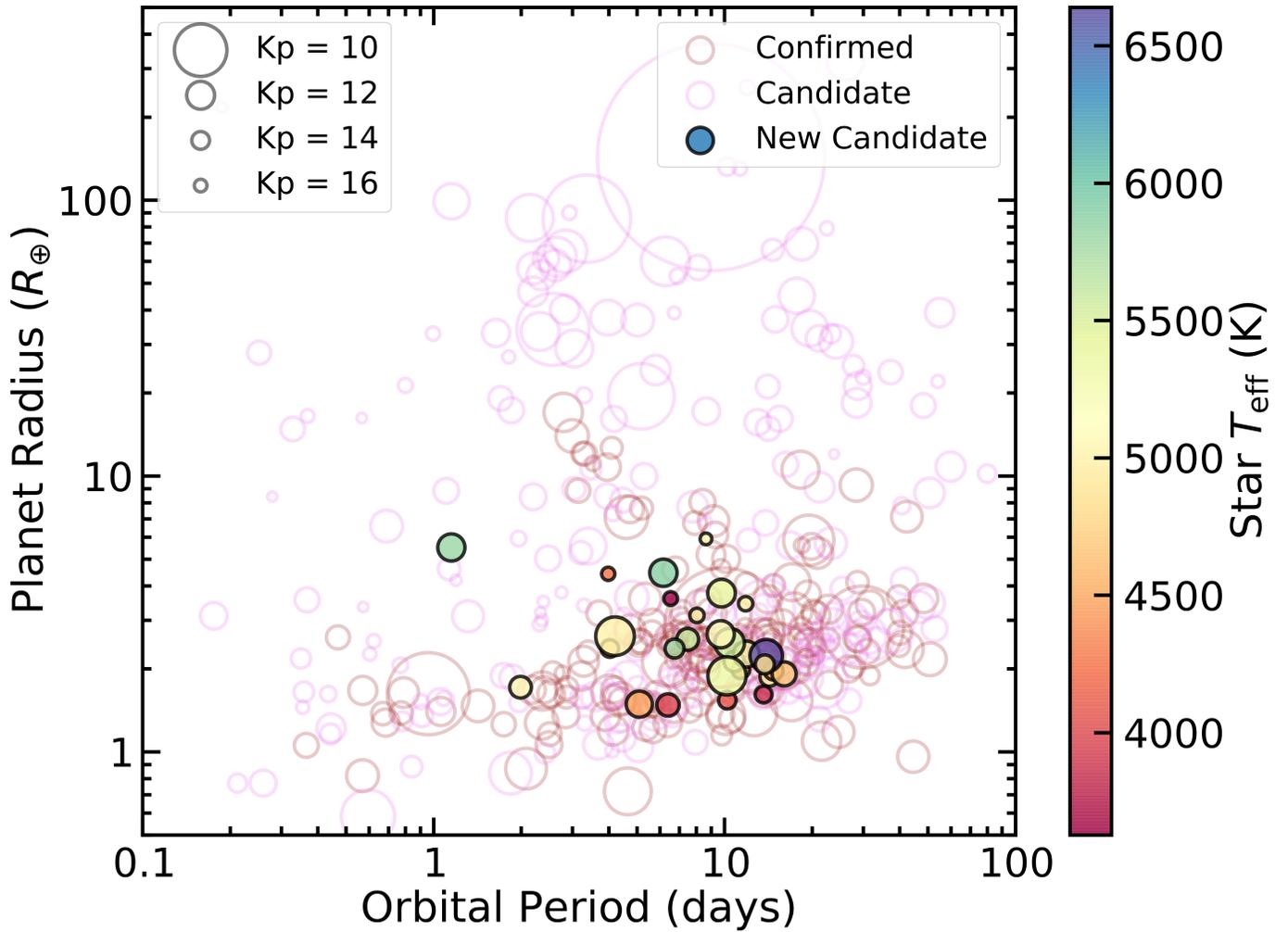}
\end{center}
\caption{Planet orbital period versus planet radius for our new candidate sample (filled circles) compared to previous \emph{K2} planet candidates (violet open circles) and confirmed planets (brown open circles) from the NASA Exoplanet Archive. Circle size indicates \emph{Kepler} bandpass magnitude, and color of the new candidates indicates host star effective temperature. \label{fig:population}}
\end{figure}

In Figure \ref{fig:population} we provide 28 new planet candidates that have been vetted by citizen scientists and expert astronomers. This catalog contains 9 likely rocky candidates ($R_{\rm pl}\le 2.0R_{\Earth}$) and 19 gaseous candidates ($R_{\rm pl}>2.0R_{\Earth}$). Within this list we find one multi-planet system (EPIC 246042088). These two sub-Neptune ($2.99\pm0.02R_{\Earth}$ and $3.44\pm0.02R_{\Earth}$) planets exist in a near 3:2 orbital resonance. The discovery of this multi-planet system is important in its addition to the list of known multi-planet systems within the \emph{K2} catalog, and more broadly in understanding the multiplicity distribution of the exoplanet population \citep{zin19}. The candidates on this list are anticipated to generate RV amplitudes of 0.2-18 m/s, many within the range accessible to current facilities.

\section*{Acknowledgement}

We thank all of the 21,770 volunteers who have helped classify transits via Exoplanet Explorers. The following users were unable to be contacted, but also provided early classifications: Cleaver82, DarylW, EEdiscoverer, garryway, GeorgeHolbrook, Grayzer56, Ianbourns, jgraber, krbethune, miguelgambler, mwalden, mzslashx, Or2dee2, sona25, swiese, Toncent, and willedwards45. This publication uses data generated via the Zooniverse.org platform, development of which is funded by generous support, including a Global Impact Award from Google, and by a grant from the Alfred P. Sloan Foundation. This research has made use of the Exoplanet Follow-up Observation Program website and the NASA Exoplanet Archive, which is operated by the California Institute of Technology, under contract with the National Aeronautics and Space Administration under the Exoplanet Exploration Program.

\bibliography{bib}\setlength{\itemsep}{-2mm}

\end{document}